# The conductance of porphyrin-based molecular nanowires *increases* with length

Norah Algethami, Hatef Sadeghi[*], Sara Sangtarash and Colin J Lambert[*]

Theory of Molecular Scale Transport, Physics Department, Lancaster University, Lancaster, UK

[*]h.sadeghi@lancaster.ac.uk; c.lambert@lancaster.ac.uk

**Abstract -** High electrical conductance molecular nanowires are highly desirable components for future molecular-scale circuitry, but typically molecular wires act as tunnel barriers and their conductance decays exponentially with length. Here we demonstrate that the conductance of fused-oligo-porphyrin nanowires can be either length independent or increase with length at room temperature. We show that this negative attenuation is an intrinsic property of fused-oligo-porphyrin nanowires, but its manifestation depends on the electrode material or anchor groups. This highly-desirable, non-classical behaviour signals the quantum nature of transport through such wires. It arises, because with increasing length, the tendency for electrical conductance to decay is compensated by a decrease in their HOMO-LUMO gap. Our study reveals the potential of these molecular wires as interconnects in future molecular-scale circuitry.

The search for molecular nanowires, whose electrical conductance decays slowly with length has been subject to many studies in the last couple of decades[1–6]. Single-molecule wires typically act as tunnel barriers and their conductance $G$ decays exponentially by molecular length[7,8] $L$ as $G = A\, e^{-\beta L}$ where $A$ is pre-factor and $\beta$ is the decay (attenuation) factor. Molecular wires usually possess a high beta factor, which limits their potential as interconnects in future molecular-scale circuitry. For example, measured room-temperature values of $\beta$ range from 2.0 - 3.4 $nm^{-1}$ for OPEs[9], 3.3 $nm^{-1}$ for OAEs[10], 1.7 - 1.8 $nm^{-1}$ for OPVs[11], 4.9 $nm^{-1}$ for acenes[12], 1.7 - 3.1 $nm^{-1}$ for oligoynes[11,13] and 8.4 $nm^{-1}$ for alkanes[14] depending on their precise anchor groups to gold electrodes.

The aim of the present paper is to identify molecular wires with vanishing or even a negative value of $\beta$, motivated by measurements of molecular wires based on porphyrin derivatives[15–20], which exhibit exceptionally low attenuation factors, due to their highly conjugated electronic structure. For example, scanning tunneling microscope (STM) measurements using a gold tip and substrate revealed that molecular wires formed from porphyrin units connected to each other through acetylene linkers exhibit a low attenuation factor of $\beta$ =0.4 $nm^{-1}$ with both pyridyl and thiol anchors[2,21] and fused-oligo-porphyrin wires with pyridyl anchors[22] exhibited an even lower value of $\beta = 0.2$ $nm^{-1}$. In what follows, we demonstrate that by employing different anchors, this fascinating family of molecular wires can exhibit vanishing or negative attenuation factors.

Here we compute the electrical conductance of the highly conjugated porphyrin wires shown in figure 1, in which neighbouring porphyrins are fused to each other via three single bonds (shown in red in fig. 1). We systematically examined fused-oligo-porphyrin (FOP) wires with different lengths connected to different electrodes with different anchors and consistently found that the conductance of these fused-oligo-porphyrin (FOP) wires can increase with length and that they possess a negative attenuation factor. This is first time that negative $β$–factor wires have been identified and is significant, because these wires are stable and therefore ideal candidates for low-conductance interconnects. To demonstrate that this result is generic and occurs for different electrode materials and anchor groups, we study quantum transport through FOPs (fig. 1a) with three different lengths (fig. 1b,c,d) sandwiched between either gold electrodes[23,24] with thiol or acetylene anchors. We also study FOPs between graphene electrodes[17,25,26] with either direct carbon-carbon bonds to the edges of the graphene or non-specific, physisorbed coupling to the graphene.

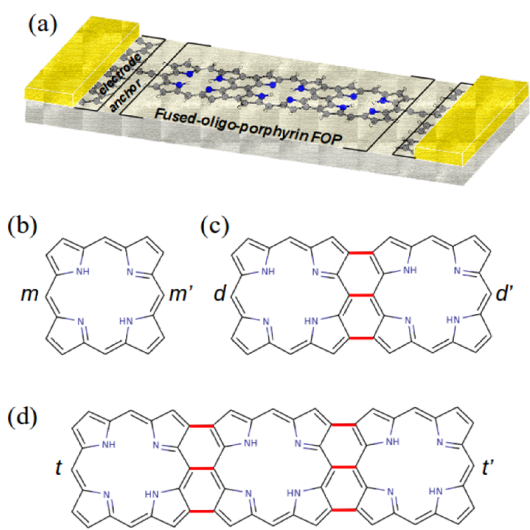

Fig. 1: A schematic of a generic molecular junction and fused-oligo-porphyrin (FOP) monomer, dimer and trimer molecular wires. (a) Shows the schematic of a generic molecular junction containing a fused porphyrin trimer. (b) a porphyrin monomer connected to electrodes from *m* and *m'* connection points (c) A fused porphyrin dimer, comprising two monomers connected to each other through three single bonds (red bonds) and connected to electrodes from *d* and *d'* connection points and (d) A fused porphyrin trimer connected to electrodes from *t* and *t'* connection points.

Figure 1 shows the molecular structure of a porphyrin monomer, a fused dimer and a fused trimer, in which two or three porphyrins are connected to each other through three single bonds (shown by red lines in fig. 1c, 1d). We first consider molecular junctions in which the carbon atoms labelled (*m,m'*), (*d,d'*) and (*t,t'*) respectively are connected to electrodes via acetylene linkers (see SI for the molecular structure of junctions). Figure 2a shows an example of the junction with graphene electrodes (see fig. S1a-c in the SI for the detailed molecular structure) where the porphyrin wires are connected to the edges of rectangular shaped graphene electrodes with periodic boundary conditions in the transverse direction. To calculate the room temperature electrical conductance $G$, we calculate the electron transmission coefficient $T(E)$ using the Gollum transport code[27] combined with the material specific mean field Hamiltonian obtained from SIESTA implementation of density functional theory (DFT)[28] and then evaluate $G$ using the Landauer formula (see methods). Results for the monomer, dimer and trimer attached to graphene electrodes (see figure 2a) are shown in figure 2b.

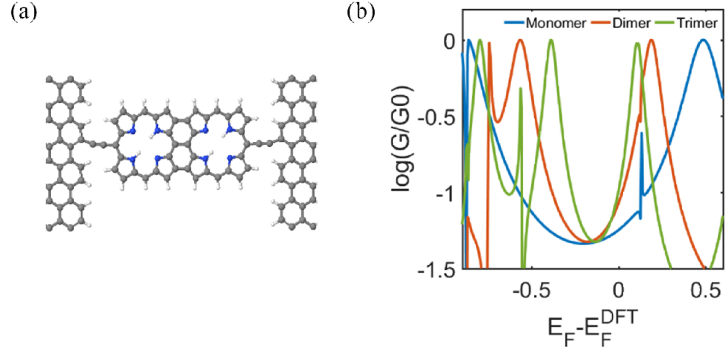

Fig. 2: Transport through monomer, dimer and trimer molecular wires attached to two graphene electrodes. (a) A fused porphyrin molecular wire connected to graphene electrodes via acetylene linkers. (b) the room temperature electrical conductance for the porphyrin monomer (blue curve), porphyrin dimer (red curve) and porphyrin trimer (green curve) as a function of the electrode Fermi energy $E_F$, in units of the conductance quantum $G_0$ = 77 micro siemens.

For these highly-conjugated wires, the energy level spacing decreases as their size increases. Therefore, the energy gap between the highest occupied molecular orbital (HOMO) and lowest unoccupied molecular orbital (LUMO) of the dimer is smaller than that of the monomer and in turn, the HOMO-LUMO (HL) gap of the trimer is smaller than that of the dimer. This behaviour is reflected in the conductance resonances of figure 2b, which are furthest apart for the monomer (blue curve) and closest together for the trimer (green curve). This can be understood by starting from a chain of N isolated monomers. Since each monomer has a HOMO energy $E_H^0$ and a LUMO energy $E_L^0$, the isolated chain has N-fold degenerate HOMO and N-fold degenerate LUMO. When the monomers are coupled together to form a fused wire, the degeneracies are lifted, to yield a HOMO, N-tuplet with molecular orbital energies $E_H^1 < E_H^2 < \cdots < E_H^0 \ldots < E_H^N$ and a LUMO, N-tuplet $E_L^1 < E_L^2 < \cdots < E_L^0 \ldots < E_L^N$. Consequently the new HL gap $\Delta(N) = E_L^1 - E_H^N$ is lower in energy than that of the monomer.

Figure 2b shows the electrical conductance as a function of the electrode Fermi energy $E_F$, plotted relative to the value $E_F^{DFT}$ predicted by DFT for pristine electrodes. The precise value of the electrode Fermi energy $E_F$ can depend on many environmental factors, but unless the molecular energy levels are shifted by an electrostatic or electrochemical gate, it always lies within the H-L gap of the contacted molecule. If $E_F - E_F^{DFT}$ is approximately -0.18 eV, then all three curves in figure 2b coincide and the monomer, dimer and trimer will possess the same conductance. For any other value within the HL gap (ie between the resonant peaks in the range -0.4 eV to +0.1 eV) the conductance of the trimer exceeds that of the dimer, which in turn exceeds that of the monomer. Consequently we predict that β is negative or zero.

To demonstrate that negative values of β are a generic feature of FOP molecular wires and occur for different choices of electrode or anchor groups, we calculated their electrical conductances when connected to gold electrodes through thiol anchors (fig. 3a). We also computed their conductances when coupled to graphene electrodes without a conventional anchor group (fig. 3c). For these molecular junctions, figures 3b, d show the corresponding electrical conductance. For the gold junctions with thiol anchors figure 3b shows that for the thiol-anchored wires, if $E_F - E_F^{DFT}$ is lower than the mid-gap (0.18 eV) of the trimer, $\beta$ is zero or slightly positive, otherwise $\beta$ is negative.

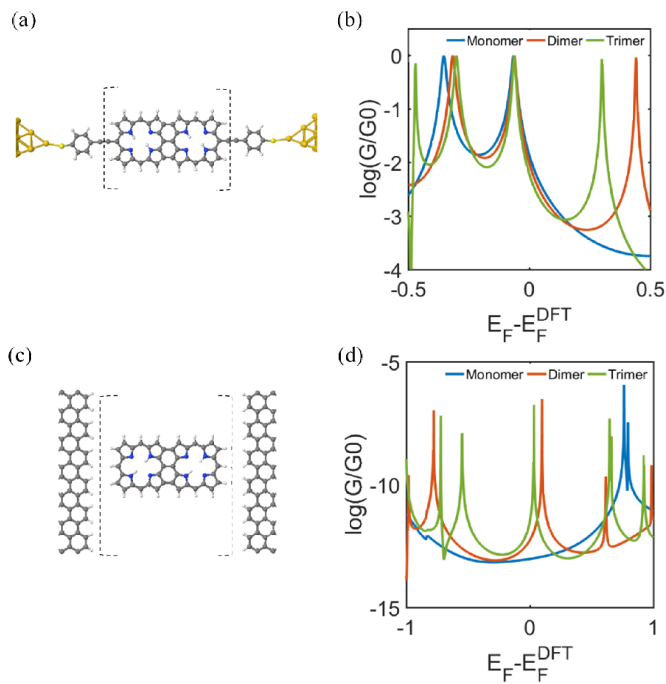

Fig. 3: Transport through fused monomer, dimer and trimer porphyrin wires sandwiched between two graphene or gold electrodes. (see figure S2 and S3 in the SI for molecular structure of all junctions). (a) A gold/dimer/gold junction with thiol anchors. (b) The electrical conductances of the gold/monomer, dimer or trimer/gold junctions with thiol anchors. The distance between the molecules with gold electrodes and carbon (sulfur) atom is 0.214 nm (0.26 nm). (c) graphene/monomer, dimer or trimer/graphene junctions without specific anchoring to the graphene. The distance between FOP and graphene electrodes are 0.3 nm. (d) The electrical conductances graphene/monomer, dimer or trimer/graphene junctions without specific anchoring to the graphene.

In the graphene junctions without specific anchoring (figs. 3 f), where the overall conductance is low due to the weak physisorbed nature the coupling to the electrodes, the electrical conductances of FOPs within the HL gap of the trimer are again found to increase with length. This unconventional negative beta factor is clearly independent of the connection point to the electrodes, because in the junctions of figure 3c (see fig. S3 in the SI for details of structure) there is no specific connection point between the electrode surfaces and the molecules. The results of figures 2 and 3 demonstrate that low or negative $\beta$ factors are a common feature of fused oligoporphyrins and occur for different modes of anchoring to electrodes.

To clarify why the conductance increases with length, we constructed a simple tight-binding model, in which a single *p* orbital per atom interacts with nearest neighbour orbitals only. The energy origin is chosen such that all on-site energies are zero except for nitrogens (see method) and the energy scale is chosen such that all intra-porphyrin nearest-neighbour couplings are set to $\gamma=-1$. We calculated the transmission function $T(E)$ between two ends of the wires e.g. (with contact atoms $(m,m')$, $(d,d')$ and $(t,t')$ for the monomer, dimer and trimer respectively, as shown in fig. 1). We then examined the effect of varying the coupling parameter α between neighbouring porphyrin units (shown by red bonds in fig. 1c,d). The different curves in figure 4 show that for a value α = -0.65γ, the curves overlap and for more negative values of α, the transmission coefficient increases with length for energies within the HL gap of the trimer (fig. 4a), in agreement to the above DFT results. To demonstrate that the decrease in the HL gap is due to a splitting of the HOMO and LUMO degeneracies, figure 4b shows the transmission curves of the trimer over a larger range of energy, for a series of values of the coupling α. For small α, the HOMO and LUMO are each almost triply degenerate and as the magnitude of α increases, the degeneracy is increasingly lifted, leading to a reduction in the HL gap.

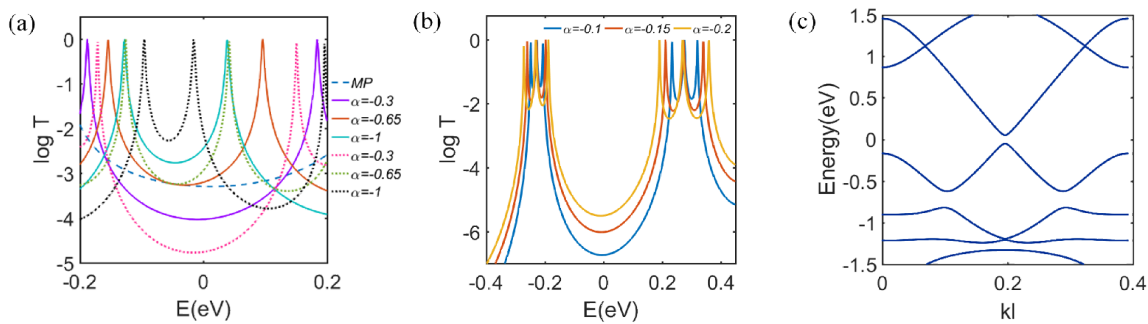

Fig. 4: The transmission coefficient for three connections point (m,m'), (d,d') and (t,t') shown in fig. 1b, 1c and 1d respectively, obtained using a simple tight binding TB model of FOP junctions. (a) The dash line curve shows the transmission coefficients for the monomer. The solid and dotted lines show the transmission coefficient for the dimer and trimer respectively. The solid red and dotted green curves show the transmission coefficient for the dimer and trimer when α=-0.65. (b) The transmission coefficient of the trimer for values of α = -0.1, -0.15, -0.2. (c) Band Structure of fused porphyrin nanoribbon.

For α=0.65γ, figure S7 of the SI shows that this increase in conductance with length persists even if the number of fused porphyrin units increased to 4, 5 and 6 units. On the other hand, figure 4c shows that the band structure of an infinitely-periodic fused porphyrin wire, calculated using density functional theory, possesses a small energy gap of about ~100meV. Therefore, fused porphyrin ribbons are narrow-gap semiconductors, meaning that eventually the conductance will begin to decrease with length. In practice, this decrease is likely to be slower than exponential, because at room temperature and large enough length scales, inelastic scattering will become significant and a cross-over from phase-coherent tunnelling to incoherent hopping will occur[10]. For comparison, figure S6 of the SI shows the transmission curves for butadiyne-linked porphyrin monomer, dimer and trimer molecular wires, for which the attenuation factor $\beta$ is clearly positive for a wide range of energies within the HL gap of the trimer in agreement with the reported measured values[21]. The fact that fused porphyrin ribbons are narrow-gap semiconductors means that for a finite oligomer, when electrons tunnel through the gap there will be contributions to the transmission coefficient from both the HOMO and the LUMO bands. Figure S9 of the SI shows that the qualitative features of figure 4a and figure 2 can be obtained by summing these two contributions.

The tight-binding results of figure 4 and the DFT results with a non-specific anchor (figure 3) suggest that a negative beta factor is a generic feature of the fused porphyrin core, provided the centres of the HOMO-LUMO gaps of the monomer, dimer and trimer are coincident. However whether or not it is measured experimentally depends on level shifts of molecular orbitals after attaching to the electrodes. This is illustrated by the calculations shown in figure S10 in the SI using direct C-Au covalent anchoring to gold electrodes, where the HOMOs of the monomer, dimer and trimer coincide and therefore the centres of their HOMO-LUMO gaps are not coincident. This spoils the generic trend and leads to a positive beta factor.

In summary, we have demonstrated that the electrical conductance of fused oligo porphyrin molecular wires can either *increase* with increasing length or be length independent in junctions formed with graphene electrodes. This is due to alignment of the middle of the HOMO-LUMO gap of the molecules with the Fermi energy of the graphene electrodes. In addition, we show that in junctions formed with gold electrodes, this generic feature is anchor group dependent. This negative attenuation factor is due to the quantum nature of electron transport through such wires and arises from the narrowing of the HOMO-LUMO gap as the length of the oligomers increases.

**Computational Methods**

The Hamiltonian of the structures described in this paper was obtained using DFT (as described below) or constructed from a simple tight-binding approximation with a single orbital per atom of site energy $\varepsilon = 0$ and $-0.5/\gamma$ for carbon and nitrogen respectively and nearest-neighbour couplings $\gamma = -1$ for both C-C and C-N bonds. Single bonds connecting different FOP units in fig.1 (red bond) is $\alpha = 0.65\gamma$.

**DFT Calculation:** The geometry of each structure consisting of electrodes (either Graphene or Gold) and non-atomic porphyrin molecule was relaxed to a force tolerance of 20 meV/Å using the SIESTA[28] implementation of density functional theory (DFT), with a double-ζ polarized basis set (DZP) and generalized gradient functional approximation (GGA-PBE) for the exchange and correlation functional. A real space grid was defined with an equivalent energy cutoff 150 Ry. The k-point grid of 1x1x20 was chosen for band structure calculation.

**Transport Calculation:** The mean-field Hamiltonian obtained from the converged DFT calculation or a simple tight-binding Hamiltonian was combined with our implementation of the non-equilibrium Green's function method, Gollum[27] to calculate the phase-coherent, elastic scattering properties of each system consist of left (source) and right (drain) leads and the scattering region (molecule). The transmission coefficient $T(E)$ for electrons of energy $E$ (passing from the source to the drain) is calculated via the Landauer formula $G = G_0 \int_{-\infty}^{\infty} dE\, T(E)(-\frac{\partial f(E)}{\partial E})$ where $f(E)$ is Fermi distribution function, $G_0 = \frac{2e^2}{h}$ is the conductance quantum, $e$ is the electron charge, and $h$ is the Planck's constant.

**Acknowledgment**

H.S. acknowledges The Leverhulme Trust for Leverhulme Early Career Fellowship no. ECF-2017-186. Further support from the UK EPSRC is acknowledged, through grant nos. EP/M014452/1, EP/P027156/1, EP/N017188/1 and EP/N03337X/1.

# Supporting Information

## The conductance of porphyrin-based molecular nanowires *increases* with length

Norah Algethami, Hatef Sadeghi[*], Sara Sangtarash and Colin J Lambert[*]

Quantum Technology Centre, Physics Department, Lancaster University, Lancaster, UK

[*]*h.sadeghi@lancaster.ac.uk; c.lambert@lancaster.ac.uk*

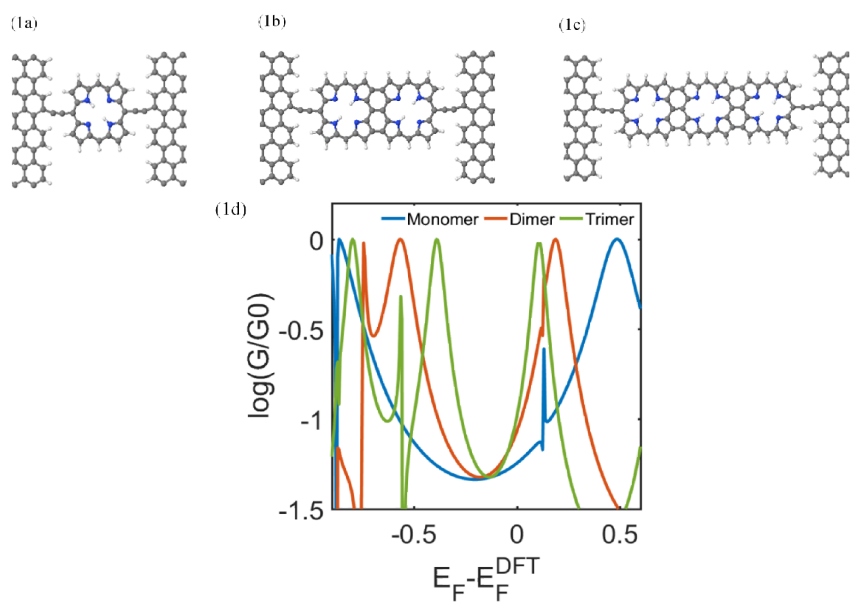

Fig. S1. Transmission coefficient obtained from DFT Hamiltonian for three types of porphyrin connect to graphene electrodes via triple bonds

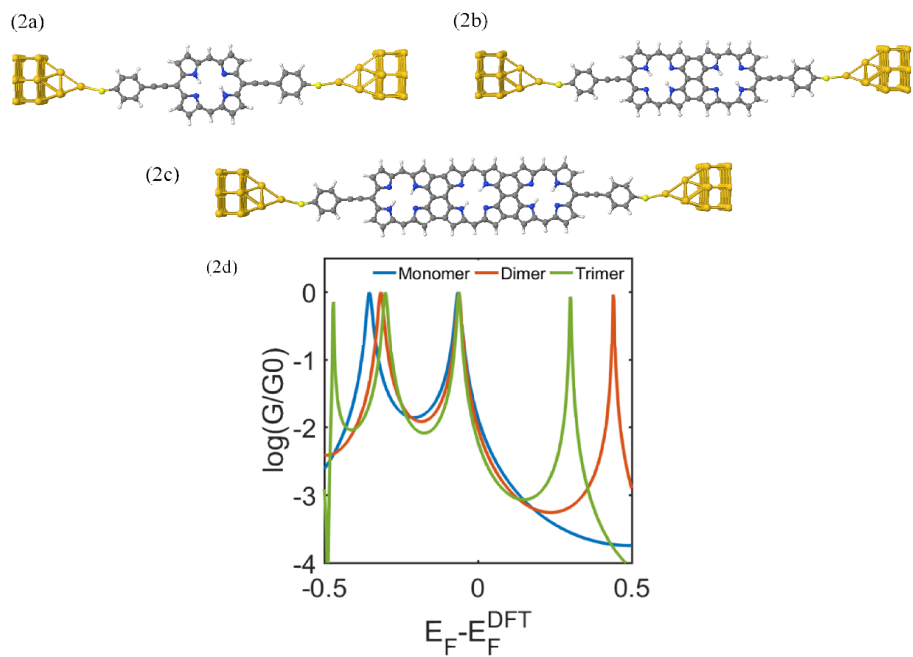

Fig. S2. Transmission coefficient obtained from DFT Hamiltonian for three types of porphyrin connected to gold electrodes via thiol-anchor

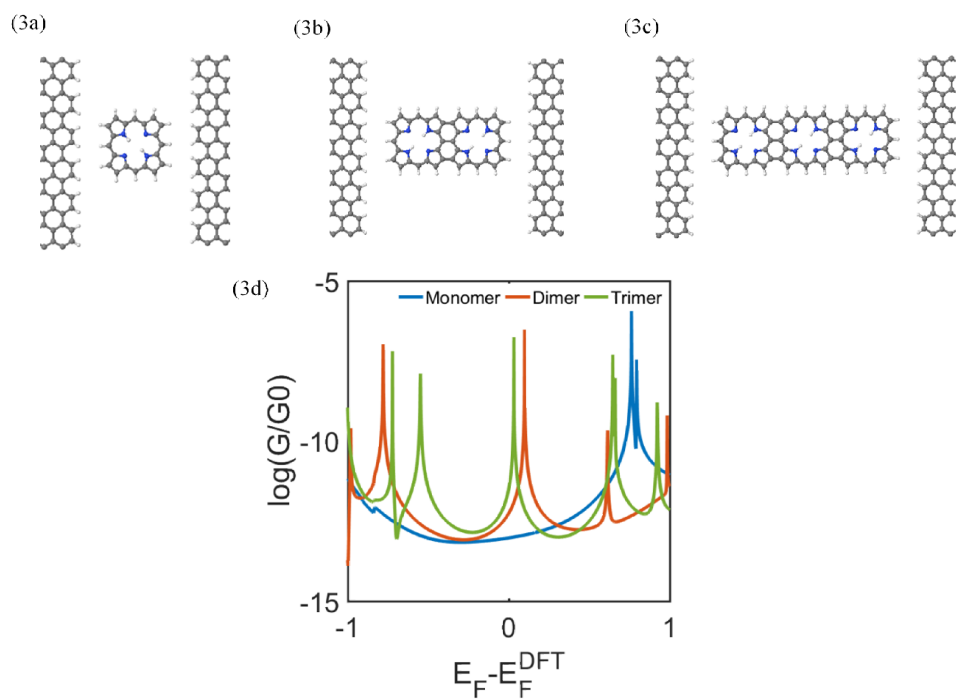

Fig. S3. Transmission coefficient obtained from DFT Hamiltonian for three types of porphyrin between graphene electrodes without specific anchor.

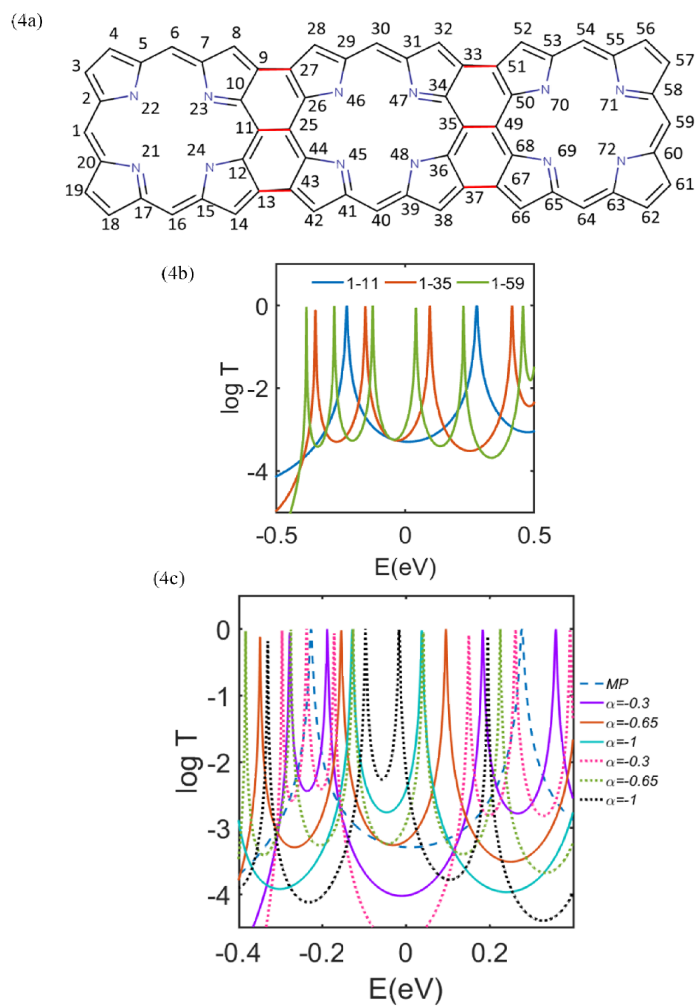

Fig. S4. Transmission coefficient for three types of porphyrin calculated using simple tight binding model

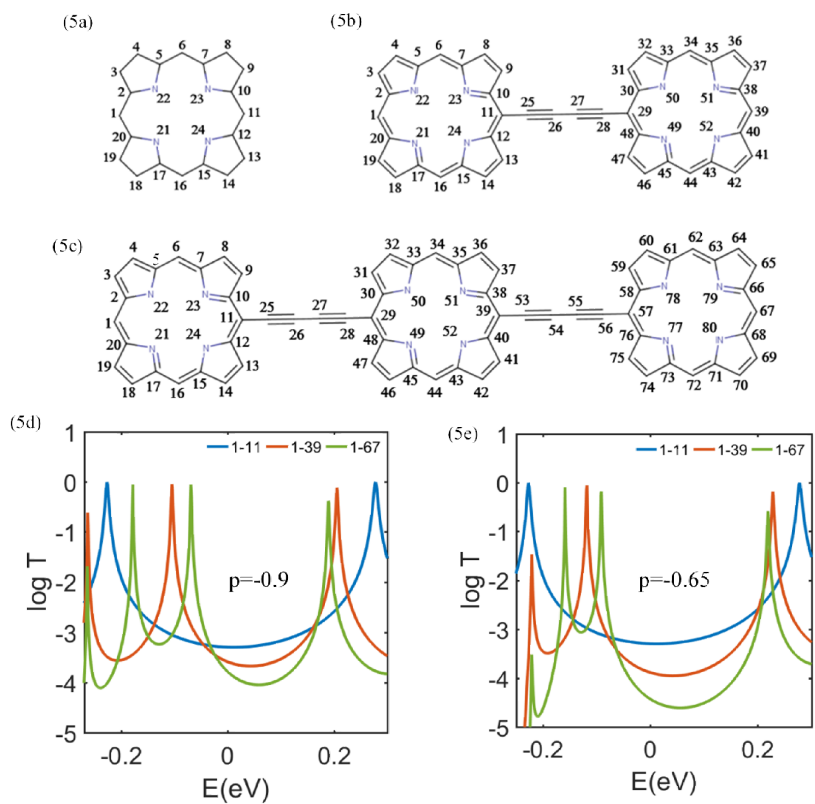

Fig. S5. (a-c) are the schematic of non-fused porphyrin monomer, dimer and trimer molecular structures. (d-e) are the transmission curves for the non-fused porphyrin calculated using simple tight binding model. Triple bond between two or three Monomer (α) integrals are chosen to be -0.9 and -0.65 in d and e, respectively.

(6a)

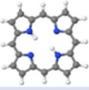

Fig. S6. Comparison between the HOMO and LUMO orbitals for (6a) fused oligo porphyrin and (6b) Non-Fused Porphyrin.

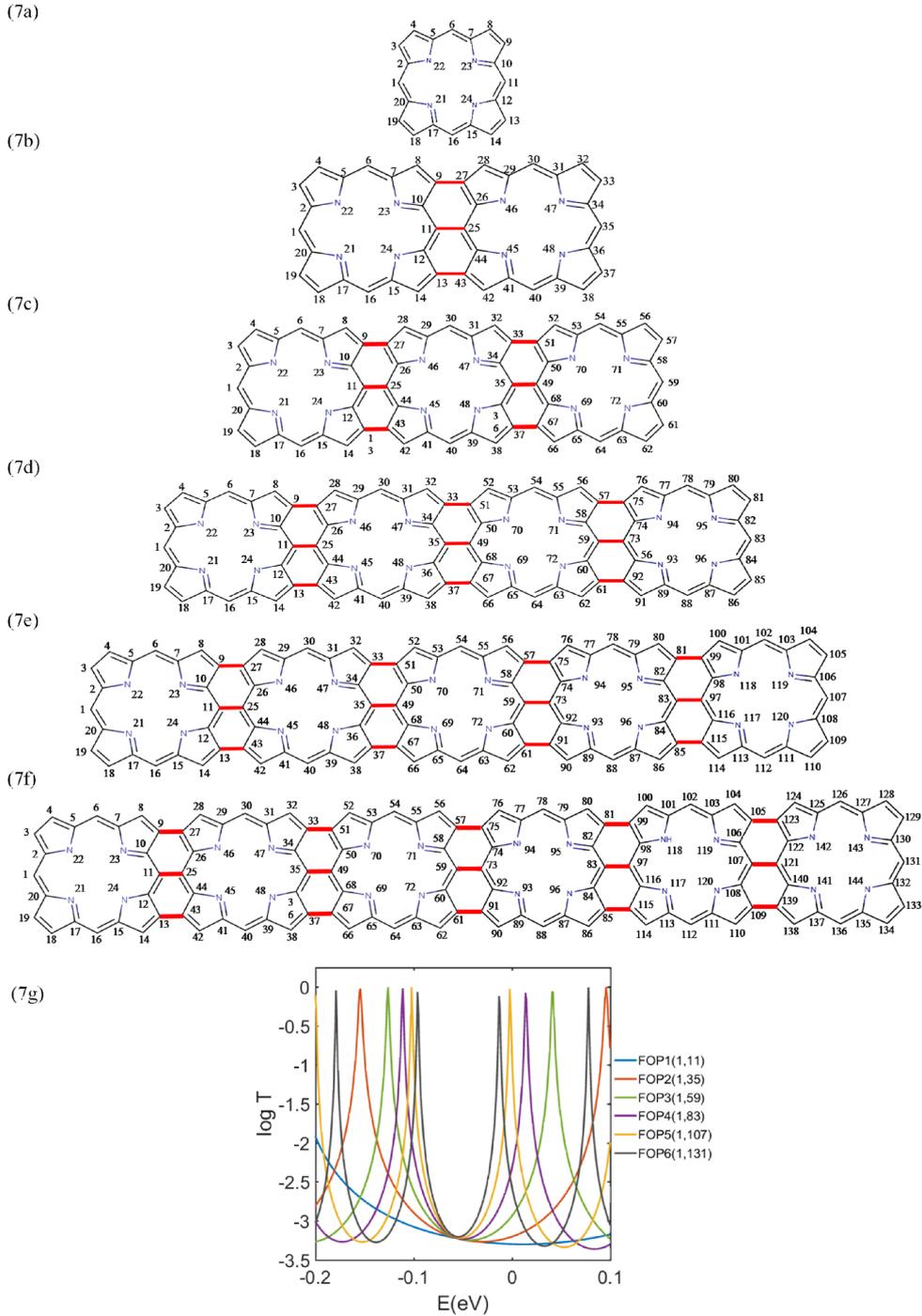

Fig. S7. Transmission coefficient for fused oligo porphyrins with different length upto 6 porphyrin units calculated using simple tight binding model. The red bonds are chosen to be $\alpha = -0.65$.

**A simple model based on coupling the frontier orbitals of a chain of monomers**

Here we note that the qualitative features of the figure 2 can be reproduced by a simple tight binding model of independent transport through the HOMOs and LUMOs. Let $T(E, n, -\gamma_L, \varepsilon_L)$ be the transmission coefficient for a chain of $n$ monomer LUMOs, with energies $\varepsilon_L$ and coupled by nearest neighbour matrix elements $-\gamma_L$. Similarly let $T(E, n, +\gamma_H, \varepsilon_H)$ be the transmission coefficient of an independent chain of monomer HOMOs, with energies $\varepsilon_H$ and coupled by nearest neighbour matrix elements $+\gamma_H$. Note that from fig 4b, since the splitting of the LUMO resonances is greater than that of the HOMO resonances, $\gamma_L > \gamma_H$. Then if we assume no interference between the HOMO and LUMO, the total transmission coefficient is

$T(E, n) = T(E, n, -\gamma_L, \varepsilon_L) + T(E, n, +\gamma_H, \varepsilon_H)$. Without loss of generality, we choose $\varepsilon_H = -\varepsilon_L$, which fixes the energy origin. As shown in figure S9, with an appropriate choice of parameters, this simple model captures the qualitative features of figure 2 and the tight-binding results of figure 4a.

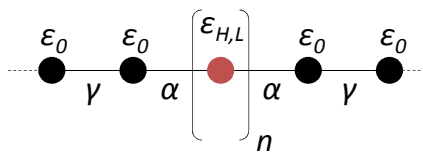

Figure S8. A tight binding (Hückel) model of 1, 2 or 3 site scattering region (depicted in red), coupled to one-dimensional leads. The scattering region represents either a chain of coupled monomer LUMOs or a chain of coupled monomer LUMOs. After calculating their separate transmission coefficients, they are simply added to give the total transmission coefficient.

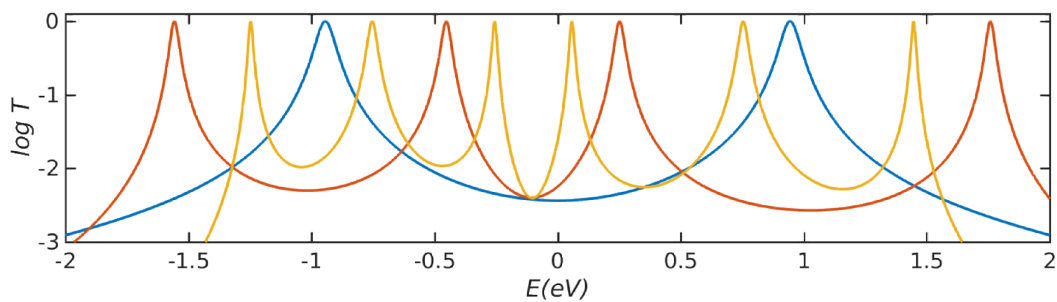

Fig. S9. Sum of transmission coefficient $T(E, n) = T(E, n, -\gamma_L, \varepsilon_L) + T(E, n, +\gamma_H, \varepsilon_H)$ through independent HOMO and LUMO levels for a monomer n=1, dimer n=2 and trimer n=3. For the monomer, $\varepsilon_L = 0.935$; for the dimer, $\varepsilon_L = 1.0$, $\gamma_L = 0.75, \gamma_H = 0.55$ and for the trimer, $\varepsilon_L = 0.75$, $\gamma_L = 0.5, \gamma_H = 0.35$. In these plots, the coupling between the molecule and the one-dimensional leads is $\alpha = -0.1$ and the leads are represented by a chain of sites with site energies $\epsilon_0 = 0$ and nearest neighbour couplings $\gamma = -1$.

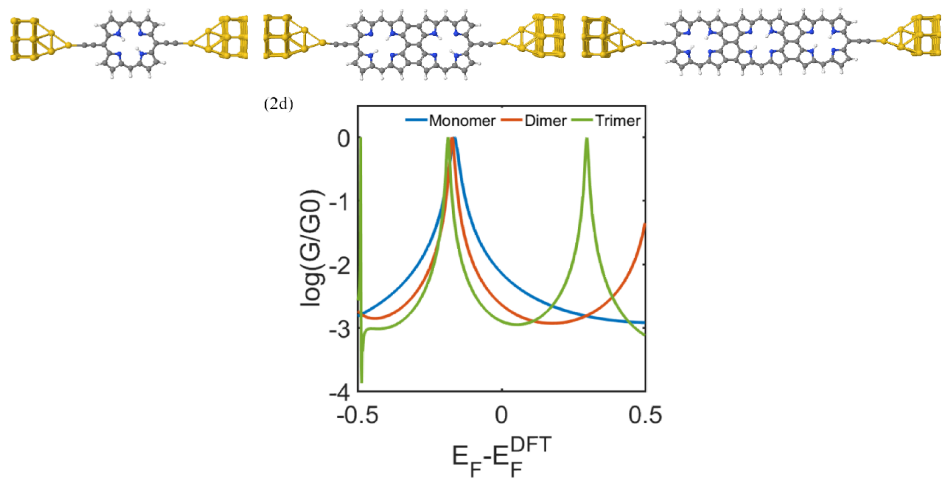

Fig. S10. Transmission coefficient obtained from DFT Hamiltonian for three types of porphyrin connected to gold electrodes through a direct Au-C bond.